\documentclass[aps,twocolumn,showpacs]{revtex4}
\makeatletter

\newcommand{\Rmnum}[1]{\expandafter\@slowromancap\romannumeral #1@}
\makeatother
\usepackage{graphicx}
\usepackage{dcolumn}
\usepackage{bm}
\usepackage{CJK}
\usepackage{amsfonts}
\usepackage{psfrag}
\usepackage{wrapfig}
\usepackage{subfigure}
\usepackage{makeidx}
\usepackage{multirow}
\usepackage{epsf}
\usepackage{hyperref}
\usepackage{amsmath}
\usepackage{cases}
\usepackage[version=3]{mhchem}
\DeclareMathOperator{\sech}{sech}

\begin{document}

\title{Several Localized Waves Induced by Linear Interference between a Nonlinear Plane Wave and Bright Solitons}
\author{Yan-Hong Qin$^{1}$, Li-Chen Zhao$^{1,2}$}\email{zhaolichen3@nwu.edu.cn}
\author{Zhan-Ying Yang$^{1,2}$}
\author{Wen-Li Yang$^{2,3}$}
\address{$^{1}$School of Physics, Northwest University, Xi'an, 710069, China}
\address{$^{2}$Shaanxi Key Laboratory for Theoretical Physics Frontiers, Xi'an, 710069, China}
\address{$^{3}$Institute of Modern Physics, Northwest University, 710069 Xian, China}

\date{\today}
\begin{abstract}

We investigate linear interference effects between a nonlinear plane wave and bright solitons, which are admitted by pair-transition coupled two-component Bose-Einstein condensate. We demonstrate the interference effects can induce several localized waves possessing distinctive wave structures, mainly including anti-dark soliton, W-shaped soliton, multi-peak soliton, Kuznetsov-Ma like breather, and multi-peak breather. Especially, the explicit conditions for them are clarified by a phase diagram based on the linear interference properties. Furthermore, the interactions between these localized waves are discussed. The detailed analysis indicate that soliton-soliton interaction induced phase shift brings the collision between these localized waves  be inelastic for soliton involving collision, and be elastic for breathers. These characters come from that the profile of solitons depend on relative phase between bright soliton and plane wave, and the profile of breathers do not depend on the relative phase. These results would motivate more discussions on linear interference between other nonlinear waves. Especially, the solitons or breathers obtained here are not related with modulational instability. The underlying reasons are discussed in detail.

\end{abstract}

\maketitle

\section{Introduction}
Nonlinear wave (NW) has been paid much attention since 1960s \cite{Zabusky,nw}. Many different families of NWs have been found, such as bright soliton \cite{BS}, nonlinear plane wave, dark soliton \cite{ds}, kink \cite{kink}, breather \cite{br}, rogue wave \cite{rw}, and vector ones of them \cite{Kevrekidis,Lakshmanan,Bludov,Zhao,Zhao1,Zhao2,Baronio,Guo}. Many of them have been observed in real
experiments or in natural environment \cite{rex1,rex2,rex3,Becker,Engels}. Nowadays, many efforts are still paid to find new families of NWs, uncover the interaction properties of them, and explain the mechanism of them. Furthermore, the applications of them are discussed in many different physical systems. For examples, vector bright soliton was used to generate Bell state in ultra-cold atoms \cite{Bells}. Very recently, bright soliton interferometry was proposed in Bose-Einstein condensate (BEC) \cite{bsinter1,bsinter2}. The interference between solitons or other NWs can  be investigated by their proper superposition \cite{nodyzhao}. Since the NWs are admitted by nonlinear systems, the linear superposition form usually fails to describe the interaction between NWs \cite{DT1,DT2,Cheng}.  Therefore, it is unusual to investigate dynamical process described by linear superposition between these NWs. In contrast to the nonlinear interference process described by nonlinear superposition \cite{nodyzhao}, the interference process described by linear superposition of NWs is seen as a linear interference process. The linear interference effects between them could induce some different wave structures, in contrast to the nonlinear interactions.

We note that it is possible to obtain some linear superposition forms of NWs in a pair-transition coupled two-component BEC system \cite{Park,Tian,BoTian,zhaoling,lingzhao}. This makes it be possible to investigate
linear interference between NWs in a real physical system. Moreover, the knowledge of linear interference in wave theory can be used to explain the dynamical process clearly. Therefore, we study on linear interference between NWs in a pair-transition coupled two-component BEC system. Among the well-known NWs, bright soliton and nonlinear plane wave are two of the simplest cases. For simplicity and without losing generality, we would like to discuss the simplest case here to uncover the essential properties of linear interference between NWs.

\begin{figure}[htb]
\begin{center}
\includegraphics[height=65mm,width=80mm]{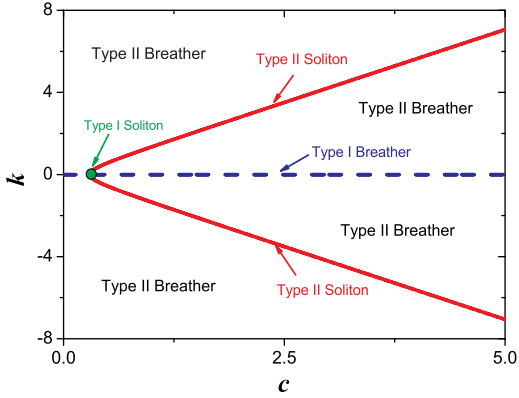}
\end{center}
\caption{A phase diagram for localized waves induced by the linear interference between one bright soliton and a nonlinear plane wave. The phase diagram is shown in wave vector $k$ and amplitude $c$ of the plane wave, with fixed soliton's amplitude and velocity. It is seen that there are two types of soliton excitations and two types of breather excitations. The explicit conditions for them are clarified by the interference period factors. The blue dashed line corresponds to the spatial interference period is infinity, and the red solid line corresponds to the temporal interference period is infinity. The parameters are
$a_1=0$, $b_1=0.21$.}\label{Fig1}
\end{figure}

In this paper, we demonstrate the linear interference effects between bright soliton and nonlinear plane wave can induce several different localized waves possessing distinctive wave structures, mainly including anti-dark soliton, W-shaped soliton, multi-peak soliton, Kuznetsov-Ma (K-M) like breather, and multi-peak breather. It is emphasized that these localized waves are different from anti-dark soliton, W-shaped soliton, multi-peak soliton and K-M breather reported before in other nonlinear systems \cite{Ws,AD,Liuchong,K-Mb}. Explicitly, previously reported ones admitted nonlinear superposition form and more complicated expressions. The existence conditions for them are clarified clearly based on the spatial-temporal interference factors (summarized in Fig. \ref{Fig1}). The relative phase between bright soliton and plane wave are also discussed in detail for the different four cases. Furthermore, the interactions between these localized waves are discussed. The detailed analysis indicate that soliton-soliton interaction induced phase shift brings the collision between these localized waves can be inelastic for soliton involving collision, and be elastic for breathers. The underlying reasons for these characters are discussed.

The outline of this paper is as follows. In Sec. \Rmnum{2}, we present the physical model and linear superposition form between bright solitons and a nonlinear plane wave. In Sec.  \Rmnum{3}, we analyze the distinctive structures of several nonlinear localized waves induced by linear interference effects between one soliton and a plane wave. A phase diagram for these localized waves are presented. In Sec.  \Rmnum{4}, we discuss the interactions between these nonlinear localized waves. The collision between them can be elastic and inelastic. The explicit conditions for elastic or inelastic collision are clarified. The underlying reason for these collision characters is explained well. In Sec.  \Rmnum{5}, we explain why soliton or breather obtained here do not involve with modulational instability. Furthermore, we suggest a possible way to clarify which dispersion form is chosen for a weak perturbation, when the perturbations admit more than one dispersion forms. The conclusion and discussion are made in Sec. \Rmnum{6}.

\begin{figure}[htb]
\begin{center}
\includegraphics[height=68mm,width=85mm]{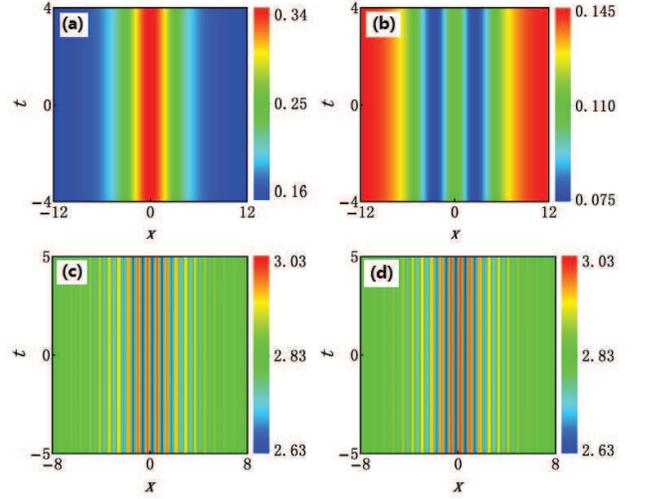}
\end{center}
\caption{The densities distribution of two types of soliton excitations. (a) and (b) for type I soliton, which correspond to component $q_{11}$ and component $q_{12}$ respectively. (c) and (d) for type II soliton, which correspond to component $q_{11}$ and component $q_{12}$ respectively. The parameters are $a_1=0$, $b_1=0.21$, $k=0$, $c=\sqrt{2}b_1$, $\phi=\frac{5\pi}{6}$ for (a) and (b); $a_1=0$, $b_1=0.21$, $k=8$, $c=\sqrt{32+2b_1^2}$, $\phi=\frac{\pi}{2}$ for (c) and (d).}\label{Fig2}
\end{figure}

\section{The physical model and linear superposition form solution}
One-dimensional two-component BEC system with particle transition can be described by the Hamiltonian
$\hat{H}=\sum_j [-\frac{\hbar^2}{2m} \hat{q}_j^\dag \partial_x^2 \hat{q}_j + \frac{g_{j,j}}{2}\hat{n}_j \hat{n}_j+
g_{j,3-j}\hat{n}_j \hat{n}_{3-j}+J_1 (\hat{q}_j^\dag \hat{q}_{3-j}+\hat{q}_{3-j}^\dag \hat{q}_j)+\frac{J_2}{2} (\hat{q}_j^\dag \hat{q}_j^\dag \hat{q}_{3-j}\hat{q}_{3-j}+\hat{q}_{3-j}^\dag \hat{q}_{3-j}^\dag %
\hat{q}_j \hat{q}_j)]$ where $n_j=\hat{q}_j^\dag \hat{q}_j$ is the particle number operator, the symbol $^{\dag}$ represents the Hermite conjugation. $g_{i,i} $ and $g_{3-i,i}$ $(i=1,2)$ are the intra and external interactions between atoms. $J_1$ and $J_2$ denote single particle and pair particles transition coupling strength separately \cite{Fischer1,Fischer2}. In most studies,  $J_{1,2}$ are set to be zero usually because it was  believed that the presence of tunneling makes the systems become non-integrable \cite{Baronio,Zhao,Lakshmanan,Bludov}. Recent experimental results in a double-well Bose-Einstein condensate suggested that pair-tunneling can become dominant with strong interaction between atoms \cite{pair,Meyer}. Therefore, we consider that the case for second-order transition is dominant, namely, $J_1=0$ and $J_2\neq 0 $.  We find integrable CNLS-p can be derived from the Hamiltonian with $g_{j,3-j}=2 g_{j,j}=2 J_2$. It is convenient to set $g_{j,j}=-\sigma$ ($\sigma=\pm 1$
correspond to attractive or repulsive interactions between atoms) without losing generality,  since there is a trivial scalar transformation for different values.

\begin{figure}[htb]
\begin{center}
\subfigure[]{\includegraphics[height=36mm,width=42.5mm]{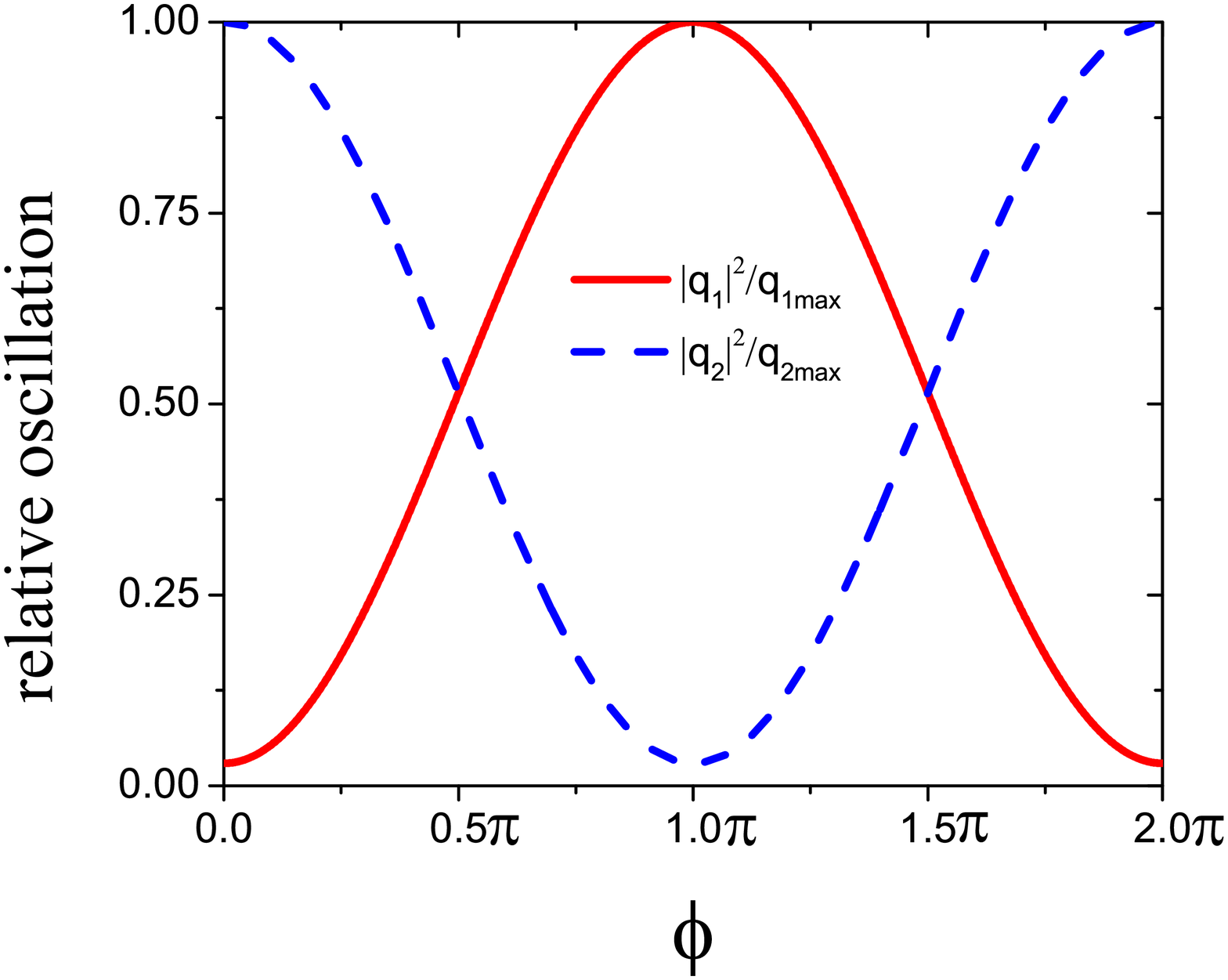}}
\hfil
\subfigure[]{\includegraphics[height=36mm,width=42.5mm]{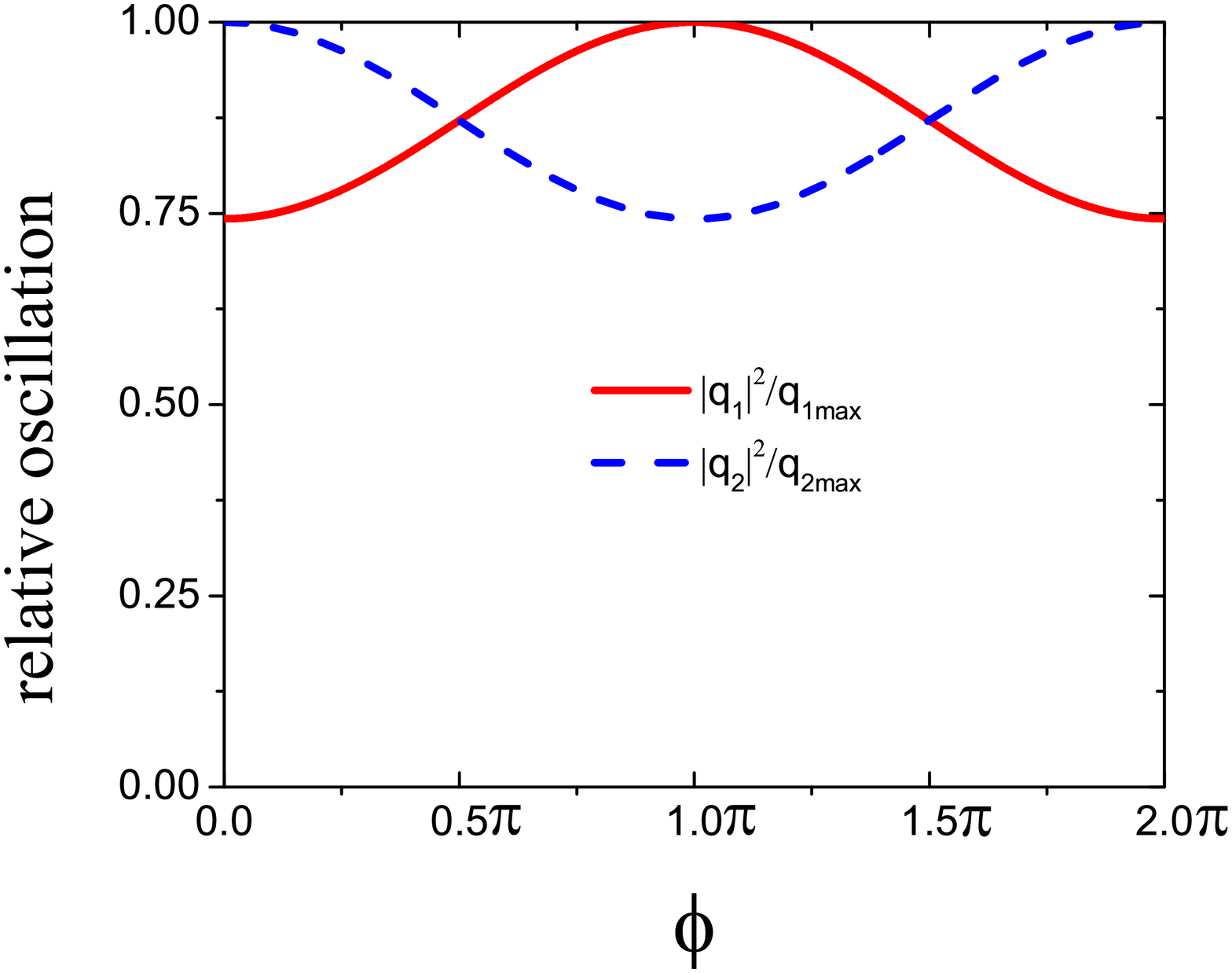}}
\end{center}
\caption{The relative density value change of two type solitons at $x=0$ with different $\phi$ value. It is shown that the profile of solitons depend on the relative phase between bright soliton and plane wave. The other parameters for (a) are identical with the ones in Fig. 2 (a) and (b), and the other parameters for (b) are identical with the ones in Fig. 2 (c) and (d). It is shown that the multi-peak soliton is much less sensitive to the relative phase than the type I soliton.}\label{Fig3}
\end{figure}

The corresponding dynamic evolution equation can be derived from the Heisenberg equation
${\rm i}\hbar(\partial \hat{q}_j/\partial t)=[\hat{q}_j,\hat{H}] $ for the field operator. Performing the mean field approximation $<\hat{q_j}>=q_j$, we can get the following integrable CNLS-p with scale dimensions $m=\hbar=1$.
\begin{equation}\label{two-mode}
    \begin{split}
      {\rm i}q_{1,t}+\frac{1}{2}q_{1,xx}+(|q_1|^2+2|q_2|^2)q_1+ q_2^2\bar{q}_1&=0, \\
      {\rm i}q_{2,t}+\frac{1}{2}q_{2,xx}+ (2|q_1|^2+|q_2|^2)q_2+q_1^2\bar{q}_2&=0,
    \end{split}
\end{equation}
where the symbol overbar represents the complex conjugation.  What needs mentioning is that the above coupled equations without the last term usually deem as non-integrable CNLS \cite{yang-benney}. However, when we add the particles transition term, the non-integrable CNLS become integrable, which was also proven by Painlev\'{e} analysis \cite{Park}.

Especially, the CNLS-p model can be transformed to two uncoupled NLS equations through a linear transformation \cite{Park,lingzhao,Lvling,Kanna,Xiang}. The solutions of Eq. (1) can be written in the form
$q_1=\frac{\psi_{1}+\psi_{2}}{2}$ and $q_2=\frac{\psi_{1}-\psi_{2}}{2}$, where $\psi_{1}$ and $\psi_{2}$ are solutions for the scalar NLS ${\rm i}\psi_{j,t}+\frac{1}{2}\psi_{j,xx}+|\psi_j|^2 \psi_j =0 $ ($j=1,2$). This is a striking character for the coupled model with particle transition terms, in contrast to the CNLS without particle transition terms \cite{Lakshmanan,Zhao,Zhao1,Zhao2,Baronio,Guo}. The scalar NLS admits many different NWs, such as bright soliton, breather, rogue wave, and nonlinear plane wave. If $\psi_1$ is one NW, and $\psi_2$ is  another NW, the linear superposition forms can be used to investigate the linear interference between two types of NWs. Here we would like to discuss the simplest case for bright solitons and a nonlinear plane wave. Namely, $\psi_1$ is a general form of nonlinear plane wave, and $\psi_2$ is bright soliton solution of the scalar NLS. The bright soliton solution can be one and multi-soliton solutions \cite{DT1,DT2}. This makes us possible to investigate linear interference between bright solitons and nonlinear plane wave systemically. Firstly, we discuss the case for one bright soliton and a nonlinear plane wave.

\section{Nonlinear localized waves induced by linear interference effects}
A general linear superposition form of one bright soliton and a nonlinear plane wave can be given as follows:
\begin{eqnarray}\label{two-mode}
q_{11}=\frac{1}{2}c e^{i \theta_{PL}}-b_1\sech[2 b_1(x+2a_1t)]e^{i \theta_{BS}}, \nonumber\\
q_{12}=\frac{1}{2}c e^{i \theta_{PL}}+b_1\sech[2 b_1(x+2a_1t)]e^{i \theta_{BS}},
\end{eqnarray}
where $c/2$ and $b_1$ denote the plane wave amplitude and soliton amplitude respectively.  $\theta_{PL}=kx+(c^2-\frac{k^2}{2})t$, and $\theta_{BS}=-2[a_1x+(a^2_1-b^2_1)t]+\phi$ describe the phase evolution of them respectively. The evolution characters are mainly determined by the phase difference between them. The parameter $\phi$ is a constant phase difference between them, which is called as relative phase here. In the following, we discuss properties of the nonlinear waves according to the two parameters $\theta_{PL}$ and $\theta_{BS}$. It is found that there are mainly four cases for interference properties according to the interference period on spatial and temporal direction. The conditions for them are summarized in Fig. \ref{Fig1}. It is obtained with variable wave vector $k$ and amplitude $c$ of the nonlinear plane wave,  and fixed soliton's amplitude and velocity. Similar phase diagram can be obtained in other cases.
\begin{figure}[htb]
\begin{center}
\includegraphics[height=68mm,width=85mm]{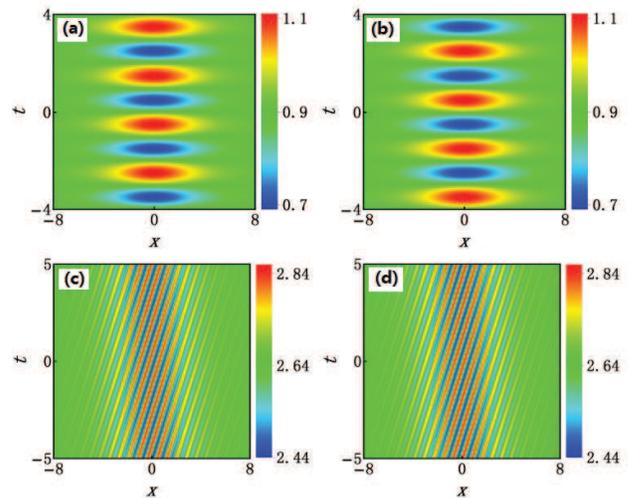}
\end{center}
\caption{The densities distribution of two types of breather excitations. (a) and (b) for type I breather, which correspond to component $q_{11}$ and component $q_{12}$ respectively. (c) and (d) for type II breather, which correspond to component $q_{11}$ and component $q_{12}$ respectively. The parameters for (a) and (b) are $a_1=0$, $b_1=0.21$, $k=0$, $c=1.8$, $\phi=\frac{\pi}{2}$. The parameters for (c) and (d) are $a_1=0$, $b_1=0.21$, $k=8$, $c=\sqrt{28 + 2 b_1^2}$, $\phi=\frac{\pi}{2}$.}\label{Fig4}
\end{figure}

Case 1: When $-2a_1=k$ and $-2(a^2_1-b^2_1)=c^2-\frac{k^2}{2}$, the temporal spatial structures of solution is presented in Fig. \ref{Fig2}(a,b). This case corresponds to the green point (type I soliton) in Fig. 1. A W-shaped soliton appears in the density distribution of the second component $q_{12}$, while an anti-dark soliton appears in the density distribution of the first component $q_{11}$. Interestingly, the soliton profile depends on the relative phase $\phi$ value. To show how the soliton profile depend on the relative phase, we define relative oscillation to be $|q_j(x=0)|^2/|q_{jmax}(x=0)|^2$ (where $|q_{jmax}(x=0)|^2$ denotes the maximum values of density value at $x=0$ vs the relative phase).   This is depicted in Fig. \ref{Fig3}(a). For $\phi= \pi /2 $, the W-shaped soliton varies to be an anti-dark soliton. It should be emphasized that the anti-dark soliton and W-shaped soliton here are distinctive from the ones obtained in other systems \cite{AD,Ws}. The superposition forms are different and their spectrums are also different, since the previously reported anti-dark soliton and W-shaped soliton all admit a nonlinear superposition form.

Case 2: When $-2a_1\neq k$ and $-2(a^2_1-b^2_1)=c^2-\frac{k^2}{2}$, densities of both components show multi-peak soliton, such structures are shown in Fig. \ref{Fig2}(c,d). This corresponds to the red solid line (type II soliton) in Fig. 1. Similarly,  the multi-peak soliton profile also depends on the relative phase $\phi$. However, the peak change with different $\phi$ in this case is much weak than the change of the case 1 (see Fig.\ref{Fig3}(b)). The visible peak number is more, and the sensitiveness on relative phase is weaker, vise versa.  Especially,  the visible peak number is determined by the soliton visible size and the wave vector of plane wave. The spatial distance between humps is $D=\frac{2 \pi}{|2a_1+k|}$, and this value should be much smaller than the soliton visible size to show multi-peak profile. Therefore, when $b_1$ is very small, $k$ is relatively larger, multi-peak structure could be observed clearly. We emphasize that the multi-peak soliton here is also distinctive from the ones obtained in \cite{Liuchong}, namely, their interference process and profiles are both distinctive.

Case 3: When $-2a_1=k$ and $-2(a^2_1-b^2_1)\neq c^2-\frac{k^2}{2}$, nonlinear localized wave is periodic on temporal direction and the behavior of density distribution is shown in Fig. \ref{Fig4}(a,b).  This corresponds to the blue dashed line (type I breather) in Fig. 1. We can see that the breathing behavior is analogous to the K-M like breather of scalar NLS \cite{K-Mb}. But the amplification rate is much smaller than the K-M breather, since there is no modulational instability gain value in this case. Interestingly, it is seen that there are humps and valleys alternately emerging in each component, and the location of hump in the component $q_{11}$ just corresponds to the valley in the component $q_{12}$. The breathing period is $T=\frac{2 \pi}{|-2(a^2_1-b^2_1)- c^2+\frac{k^2}{2}|}$.  This indicates that there are particles transition between the two components during the breathing process. The transition process admits a standard Josephson oscillation form, which is distinctive from the nonlinear one reported in \cite{zhaoling}. The profile of localized wave varies with time evolution for breather. The spatial-temporal distribution properties do not depend on the relative phase between bright soliton and plane wave anymore.

Case 4: When $-2a_1\neq k$ and $-2(a^2_1-b^2_1)\neq c^2-\frac{k^2}{2}$, we show the dynamics of the nonlinear excitation in Fig. \ref{Fig4}(c,d). This corresponds to regime except the points on the two lines in Fig. 1. The condition $-2a_1\neq k$ makes the localized wave admit multi-peak profile, and $-2(a^2_1-b^2_1)\neq c^2-\frac{k^2}{2}$ makes the localized wave breath with time evolution. To distinct from the type I breather, we call this as multi-peak breather, which corresponds to the type II breather in Fig. 1. Similar to multi-peak soliton,  the visible peak number is also determined by the soliton visible size and the wave vector of plane wave. In this case, the spatial-temporal distribution properties also do not depend on the relative phase between bright soliton and plane wave anymore.

Based on the above discussions on soliton types and profiles, one can expect that the interaction between them can be investigated analytically and exactly with the aid of multi-soliton solution of NLS \cite{bscollide}. For simplicity, we just discuss the interaction between two of them.
\begin{figure}[htb]
\begin{center}
\includegraphics[height=68mm,width=85mm]{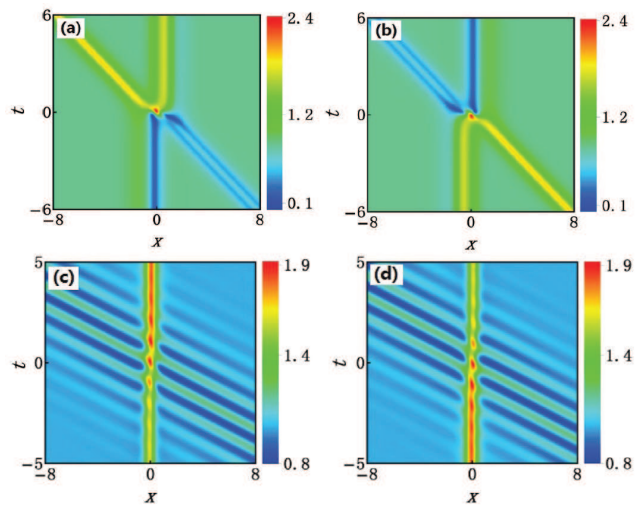}
\end{center}
\caption{ (a) and (b) The collision between type I soliton and type II soliton£¬ which  correspond to component $q_{21}$ and component $q_{22}$ respectively.  It is seen that solitons' profiles both vary greatly after collision. (c) and (d) show the collision between type I soliton and type II breather, which correspond to component $q_{21}$ and component $q_{22}$ respectively. It is seen that soliton's profile varies after collision and breather's profile is kept well. The parameters for (a) and (b) are $a_1 = 0.6, a_2 = 0, b_2 = 1,
b_1 = \sqrt{b_2^2 + a_1^2}, k = -2 a_1, c = \sqrt{2 b_2^2 + 2 a_1^2}, f_1 = f_2 = \pi/2, k_1 = k_2 = 0$. The parameters for (c) and (d) are $a_1=1$, $a_2=0$, $b_1=0.2$, $b_2=\sqrt{2}$, $c=2$, $k=0$, $f_1= f_2=\frac{\pi}{2}$, $k_1= k_2=0$.}\label{Fig5}
\end{figure}

\section{The interactions between several obtained localized waves}
The solution for two bright solitons linearly interfering with a nonlinear plane wave could be presented as follows:
\begin{eqnarray}\label{two-mode}
&&q_{21}=\frac{c}{2}e^{i \theta_{PL}}-b_1\sech[A_1]e^{iY_1}-P_2,\nonumber\\
&&q_{22}=\frac{c}{2}e^{i \theta_{PL}}+b_1\sech[A_1]e^{iY_1}+P_2,
\end{eqnarray}
where $P_2=2 b_2 \frac{X_1e^{iY_1}+X_2e^{iY_2}}{X_3}$ denotes a nonlinear superposition to the bright soliton.
$Y_{j}=f_{j}-2\alpha_{j},(j=1,2)$,
$X_{1}=-2b_1(b_1+b_2)\cosh(A_2)+2b_1^2[\cos(Y_2-Y_1)+\cosh(A_1+A_2)]\sech(A_1)+2i(a_2-a_1)b_1\sinh(A_2)$,
$X_{2}=[(a_1-a_2)^2+(b_2^2-b_1^2)]\cosh(A_1)-2i(a_2-a_1)b_1\sinh(A_1)$,
$X_{3}=2[(a_1-a_2)^2+(b_1^2+b_2^2)]\cosh(A_2)\cosh(A_1)-4b_1b_2\cos(Y_1-Y_2)-4b_1b_2\sinh(A_2)\sinh(A_1)$,
and $\alpha_{j}=a_{j}x+(a_{j}^2-b_{j}^2)t,(j=1,2)$, $\beta_{j}=b_{j}x+2a_{j}b_{j}t, (j=1,2)$, $A_{j}=k_{j}+2\beta_{j},(j=1,2)$.
The parameters $a_1$ and $a_2$ are related soliton's velocity, $b_1$ and $b_2$ determine peak value of solitons respectively. $k_1$ and $k_2$ determine the initial locations of solitons, $f_1$ and $f_2$ can be used to vary the relative phase between solitons. This solution is a linear superposition of a plane wave and a two-soliton solution, but the two-soliton solution is a nonlinear superposition of two bright solitons. When the related parameters are chosen, the solution will present us the dynamics of two localized waves directly. The localized wave type can be chosen by setting the parameters of bright soliton or plane wave background based on the phase diagram Fig. 1. The collision of arbitrary two of the above localized waves can be observed conveniently. It is found that collision between these localized waves can be inelastic for soliton involving collision, and be elastic for breathers. Nextly, we discuss on the collision between them in details.

The collision between soliton and soliton or breather can be investigated by setting the bright soliton 's parameters satisfy different conditions in Fig. 1. For examples, we show one case for type I soliton and type II soliton in Fig. \ref{Fig5} (a) and (b) by setting parameters $a_1 = 0.6, a_2 = 0, b_2 = 1, b_1 = \sqrt{b_2^2 + a_1^2}, k = -2 a_1, c = \sqrt{2 b_2^2 + 2 a_1^2}, f_1 = f_2 = \pi/2, k_1 = k_2 = 0$. The type II soliton admit multi-peak here, which is invisible in Fig. \ref{Fig5} (a) and (b). This comes from that the period is much larger than the bright soliton scale. It is shown clearly that solitons' profiles both vary after colliding process. Similarly, we can investigate collision between soliton and breather. It should be noted that it is not possible to observe collisions between type I soliton and type I breather, since their existence conditions make them admit identical velocity. From the phase diagram, we can observe the collision between type I soliton and type II breather  by choosing  $a_1=1$, $a_2=0$, $b_1=0.2$, $b_2=\sqrt{2}$, $c=2$, $k=0$.  Fig. \ref{Fig5} (c) and (d) show the collision between them, which correspond to component $q_{21}$ and component $q_{22}$ respectively. It is seen that soliton's profile varies after collision and breather's profile is kept well.

\begin{figure}[htb]
\begin{center}
\includegraphics[height=68mm,width=85mm]{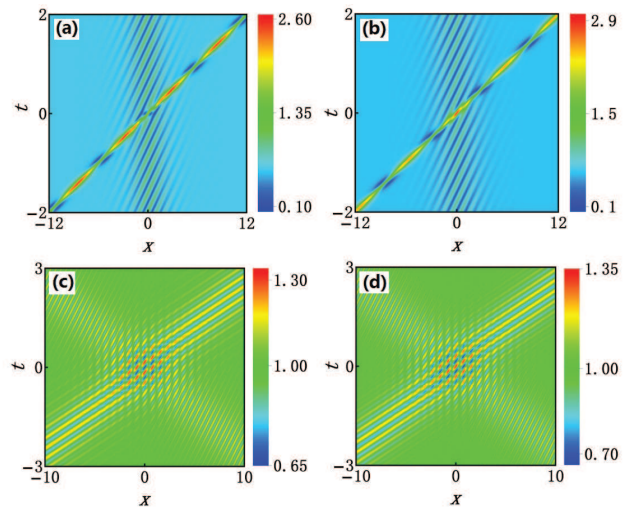}
\end{center}
\caption{The collision between breather and breather. (a) and (b) show the collision between type I breather and type II breather, which correspond to component $q_{21}$ and component $q_{22}$ respectively.  (c) and (d) show the collision between type II breather and type II breather, which correspond to component $q_{21}$ and component $q_{22}$ respectively. It is seen that breather's profile is kept well after collision. The  parameters for (a) and (b) are $a_1 = -3$, $a_2 = 0$, $b_1 =2$,  $b_2 = 0.4$, $c = 1.2$, $k = 6$, $f_1 = f_2 = 0$, $k_1 =  k_2 = 0$.
The parameters for (c) and (d) are $a_1 = 2$, $a_2 = -2$, $b_1 = 0.15$, $b_2 = 0.2$, $c = 2$, $k = 8$, $f_1 =  f_2 = 0$, $k_1 = k_2 = 0$.}\label{Fig6}
\end{figure}

The collision between breather and breather can be investigated by setting the bright soliton 's parameters satisfy different conditions in Fig. 1. For examples, type II breather and type I breather can be set by choosing $a_1 = -3$, $a_2 = 0$, $b_1 =2$,  $b_2 = 0.4$, $c = 1.2$, $k = 6$. Fig. \ref{Fig6} (a) and (b) show the collision between them, which correspond to component $q_{21}$ and component $q_{22}$ respectively.  However, it is not possible to observe the collision between type I breather and type I breather, because the existence conditions make them admit identical velocity. Type II breather and type II breather can be investigated by choosing  $a_1 = 2$, $a_2 = -2$, $b_1 = 0.15$, $b_2 = 0.2$, $c = 2$, $k = 8$.  Fig. \ref{Fig6} (c) and (d) show the collision between them, which correspond to component $q_{21}$ and component $q_{22}$ respectively. It is shown that the collision between breathers are all elastic.

Then, why the soliton's profile varies after collision? It is well known that there is a phase shift after bright soliton colliding with another one \cite{bscollide,Nguyen}. The phase shift brings relative phase changes between bright soliton and plane wave. Moreover, the soliton type localized waves' profiles depend on the relative phase. This makes the soliton's profile change. Therefore, the underlying reason for inelastic collision is soliton type localized waves depend on the relative phase between bright soliton signal and plane wave, and collision between two bright solitons can bring a phase shift on each bright soliton signal. But the breather type localized waves do not depend on the relative phase, the phase shift will not change the spatial-temporal structure of breathers.

\section{A discussion on modulational instability branches}
It should be noted that the soliton-type perturbation on the plane wave background can be both weak and strong here. We emphasize that the breathing behaviors here do not involve with modulational instability (MI). This can be seen by the above localized wave dynamics for which the perturbation amplitudes are not amplified at all even the soliton-type perturbation is very weak. MI has been shown widely to induce rogue wave or breather on a plane wave background \cite{Kibler,Dudley,Baronio1,Baronio3,zhaoling2,NMI}.  In fact, rogue wave or Akhmediev breather can also exist on the identical plane wave background in this coupled model. This can be proven by the transformation between the coupled model and standard scalar NLSE. Namely, a rogue wave solution of the coupled model can be constructed by one rogue wave solution with the identical plane wave background and one zero solution of the scalar NLSE.  This means that there are at least two dispersion relation branches on the plane wave background. We perform standard MI analysis on the plane wave background in the coupled model. We add the perturbation terms on the plane wave background, $q_1=\frac{1}{2}ce^{i[kx+(c^2-\frac{k^2}{2})t]}(1+f_{+} e^{{\rm i}\kappa (x-\Omega t)}+f_{-}^{\ast} e^{-{\rm i}\kappa (x-\Omega^{\ast} t)})$ and $q_2=\frac{1}{2}ce^{i[kx+(c^2-\frac{k^2}{2})t]}(1+g_{+} e^{{\rm i}\kappa (x-\Omega t)}+g_{-}^{\ast} e^{-{\rm i}\kappa (x-\Omega^{\ast} t)})$ (where $f_{+},f_{-},g_{+},g_{-}$ are small amplitudes of the Fourier modes). It is found that there are two dispersion relation branches, which admit one MI branch and one modulational stability (MS) branch. They can be calculated as $\Omega_{1,2}=k\pm\frac{1}{2}\sqrt{-4c^2+\kappa^2}$, and $\Omega_{3,4}=k\pm(\frac{c^2}{\kappa}+\frac{\kappa}{2})$, which correspond to MI and MS branch respectively. Since the linear stability analysis holds well for weak perturbations, there will be two possible choices for a weak perturbation on the plane wave background. Then, how to understand that the soliton-type perturbation with small amplitude evolves to be type-I or type II soliton or breather which do not involves MI characters? Namely, why the weak soliton-type perturbation choose the MS branch to evolve?

\begin{figure}[htb]
\begin{center}
\subfigure[]{\includegraphics[height=36mm,width=42.5mm]{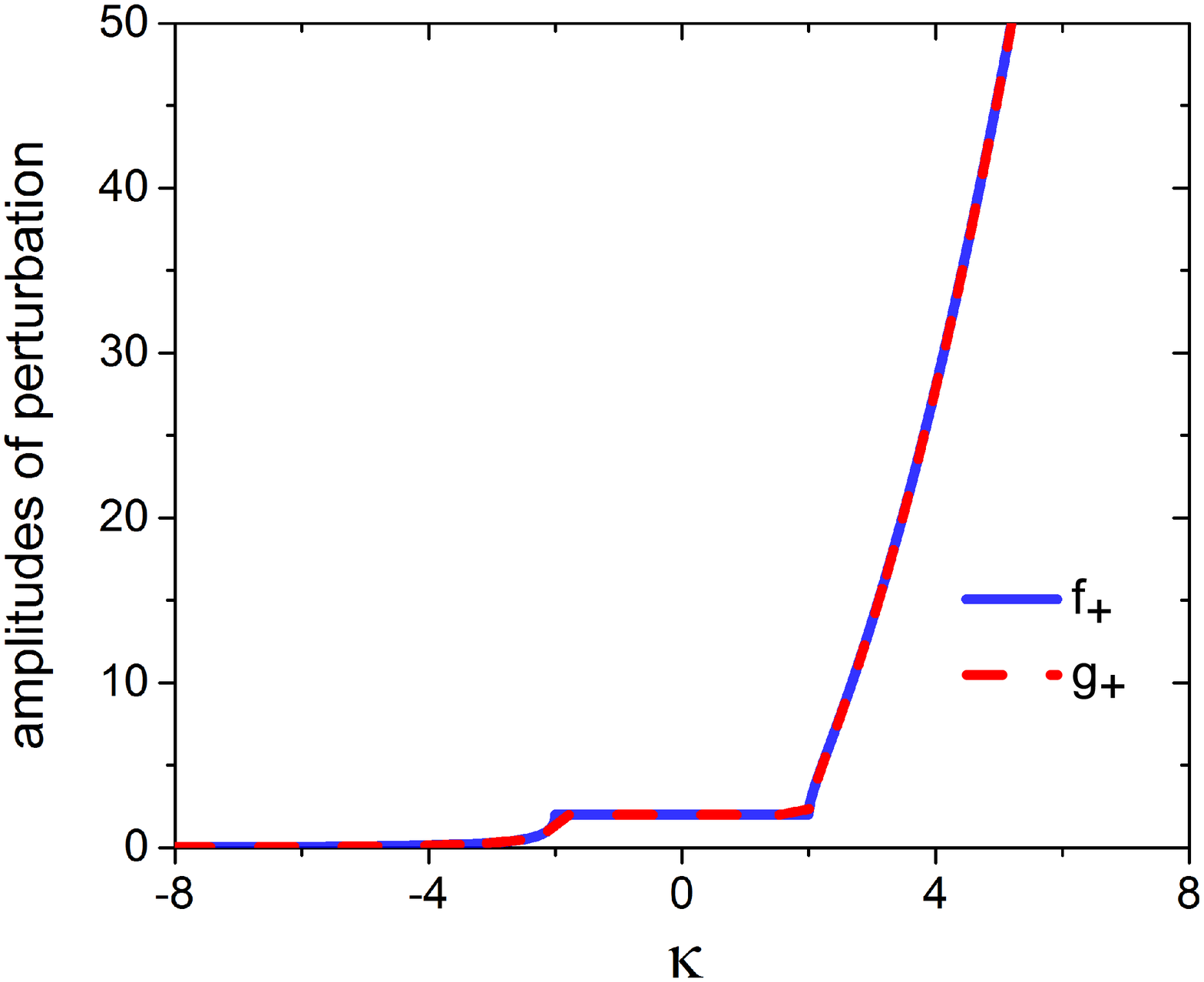}}
\hfil
\subfigure[]{\includegraphics[height=36mm,width=42.5mm]{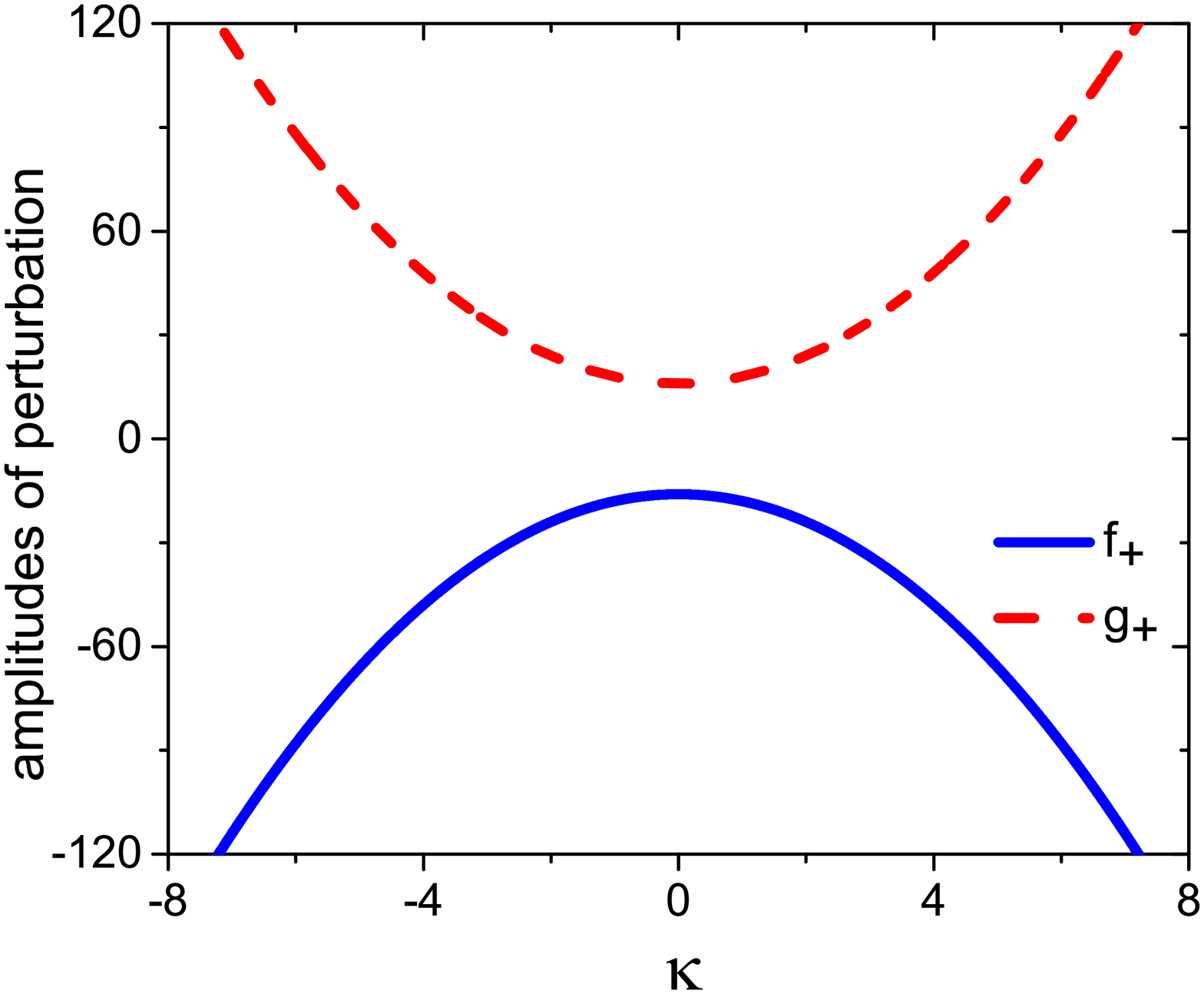}}
\end{center}
\caption{The perturbation amplitude $f_{+}$ and $g_{+}$ versus the perturbation wave vector $\kappa$. (a) and (b) correspond to  MI branch $\Omega=k+\frac{1}{2}\sqrt{-4c^2+\kappa^2}$ and  MS branch $\Omega=k+\frac{c^2}{\kappa}+\frac{\kappa}{2}$ respectively.  It is seen that $f_{+}=g_{+}$ for MI branch, $f_{+}=-g_{+}$ for MS branch. The parameter $c=1$.}\label{Fig7}
\end{figure}

We try to find the essential factors which determine choosing choice for weak perturbations, by calculating the eigenvector $(f_{+},f_{-},g_{+},g_{-})^{\mathrm{T}}$ of the linearized equations. For MI branch ($\Omega_{1,2}$), the corresponding eigenvector is $(f_{+},f_{-},g_{+},g_{-})^{\mathrm{T}}= \epsilon (-2c^2+\kappa^2\pm\kappa\sqrt{-4c^2+\kappa^2},2c^2,-2c^2+\kappa^2\pm\kappa\sqrt{-4c^2+\kappa^2},2c^2)^{\mathrm{T}} $. $\epsilon<<1$ must holds since the linear stability analysis involves linearization.  It is seen that $f_{+}=g_{+}$, $f_{-}=g_{-}$ for MI branch, namely, the perturbations form added on the backgrounds in the two components admit identical perturbation profile.  For an example, we show one case for perturbation amplitude vs perturbation wave vector $\kappa$ in Fig. 7 (a). Considering that the rogue wave solution discussed above is constructed by one rogue wave solution with the identical plane wave background and one zero solution of the scalar NLSE, we can understand that the rogue wave can exist on the background and they demonstrate MI characters, since the rogue wave solutions in the two components admit identical perturbation profile on the background and this makes the weak perturbation choose the MI branch to evolve.

For MS branch $\Omega=k+\frac{c^2}{\kappa}+\frac{\kappa}{2}$, the corresponding eigenvector can be expressed as $(f_{+},f_{-},g_{+},g_{-})^{\mathrm{T}}= \epsilon (-2(8c^2+\kappa^2),0,2(8c^2+\kappa^2),0)^{\mathrm{T}}$.
For $\Omega=k-(\frac{c^2}{\kappa}+\frac{\kappa}{2})$, the corresponding eigenvector is $(f_{+},f_{-},g_{+},g_{-})^{\mathrm{T}}= \epsilon (0,-2(8c^2+\kappa^2),0,2(8c^2+\kappa^2))^{\mathrm{T}}$. We can see that the perturbation profiles are inverse in the two components for MS branch. As an example, we present the perturbation amplitude $f_{+}$ and $g_{+}$ versus the perturbation wave vector $\kappa$ for one MS branch in  Fig. \ref{Fig7}(b). The weak soliton-type perturbations presented above all satisfy the $f_{+}=-g_{+}$ for MS branch. Namely, the type-I or type II soliton or breather obtained here all admit the case that perturbation profiles on the background are inverse in the two components. This makes that the perturbations all choose MS branch to evolve. The breathers discussed above just involves linear interference mechanism and is not related with MI at all.

\section{Conclusion and discussion}
In this paper, we show that linear interference effects between a nonlinear plane wave and bright soliton, can induce anti-dark soliton, W-shaped soliton, K-M like breather, multi-peak soliton, and multi-peak breather. Their profile properties are discussed in detail. It is shown that soliton type localized waves' profile depends on relative phase between bright soliton and plane wave, but the breathers do not. Furthermore, the interactions between these localized waves are discussed. The detailed analysis indicate that soliton interaction induced phase shift brings the collision between these localized waves can be inelastic and elastic. The underlying reason for these characters are discussed. Additionally, considering the soliton profile depends on the relative phase between plane wave and bright soliton, we expect that the inelastic collision property could be used to measure the phase shift during the collision between bright solitons. Similar studies can be extended to linear interference between other nonlinear waves in other coupled systems.  Especially, the soliton or breather obtained here is proven to be not related with MI. A possible way is suggested to clarify which MI branch is chosen for a weak perturbation, when the perturbations admit more than one MI branches.

\section*{Acknowledgments}
This work is supported by National Natural Science Foundation of
China (Contact No. 11775176), and Shaanxi Province Science association of colleges and universities (Contact No. 20160216).

\end{document}